\documentclass{aa}
\usepackage{graphicx}
\usepackage{graphicx}
\usepackage{amssymb}
\usepackage{subfigure}
\usepackage{natbib}
\usepackage{tabularx}
\usepackage{longtable}
\usepackage{lscape}
\bibpunct{(}{)}{;}{a}{}{,} 
\begin{document}

   \title{Kinematics and the origin of the internal structures in HL~Tau jet (HH~151)
   \thanks{Based on observations obtained with the 6-m telescope of the Special Astrophysical Observatory of the Russian Academy of Sciences (SAO RAS). The observations were carried out with the financial support of the Ministry of Education and Science of Russian Federation (contracts no.~16.518.11.7073 and 16.552.11.7028)}}

   \author{T.A. Movsessian
          \inst{1}
          \and
          T.Yu. Magakian
          \inst{1}
          \and
           A.V. Moiseev
          \inst{2}
             }
   \institute{Byurakan Astrophysical Observatory, 0213 Aragatsotn reg.,
              Armenia\\
                \email{tigmov@bao.sci.am, tigmag@sci.am}
                     \and
            Special Astrophysical Observatory,
             N.Arkhyz, Karachaevo-Cherkesia, 369167 Russia\\
             \email{moisav@sao.ru}
              }
   \date{Received ...; / Accepted ... }

   \abstract
   { Knotty structures of Herbig-Haro jets  are common phenomena, and knowing the origin of these structures  is essential for understanding the processes of jet formation. Basically, there are two theoretical approaches:  different types of instabilities  in stationary flow, and velocity variations in the flow.}
{We investigate the structures with  different radial velocities in the knots of the HL~Tau jet as well as its unusual behaviour starting from 20\arcsec\ from the source.   Collation of radial velocity data with proper motion measurements of emission structures in the jet of HL~Tau makes it possible to understand the origin of these structures and decide on the mechanism for the formation of the knotty structures in Herbig-Haro flows. }
   {We present observations obtained with a 6 m telescope (Russia) using the SCORPIO camera with scanning Fabry-Per\'ot interferometer.  Two epochs of the observations of the HL/XZ~Tau region
   in H$\alpha$ emission (2001 and 2007) allowed us to measure proper motions for high and low radial velocity structures. }
   {The structures with low and high radial velocities in the HL~Tau jet show the same proper motion. The point where the HL~Tau jet bents to the north (it coincides with the trailing edge of so-called knot A) is stationary, i.e. does not have any perceptible proper motion and is visible in H$\alpha$ emission only.} 
   {We conclude that the high- and low- velocity structures in the HL~Tau jet represent bow-shocks and Mach disks in the internal working surfaces of episodic outflows. The bend of the jet and the brightness increase starting some distance from the source coincides with the observed stationary deflecting shock. The increase of relative surface brightness of bow-shocks could be the result of the abrupt change of the physical conditions of the ambient medium as well as the interaction of a  highly collimated flow and the side wind from XZ~Tau.}


\titlerunning{About the origin of HL~Tau jet (HH~151)}
   \authorrunning{Movsessian et al.}
    
\maketitle
%
\section{Introduction}
\label{intro}

\textbf{\textbf{}}

 The Herbig-Haro jet from \object{HL~Tau} (\object{HH~151}) was first noted by \cite{mundt83} and became one of the first discovered jets from young stellar objects (YSOs). The HH flow associated with HL~Tau has an exciting history and even today the origin of some its features is not clear. For example, it starts as the relatively faint and narrow jet from HL~Tau with a position angle of  about 51\degr\  but at a distance of about 20\arcsec\ from the star it    abruptly changes direction by about 14\degr\ and becomes much more brighter and wider. This strange behaviour was the main reason for several authors to consider this flow as an intersection of two jets: one from HL~Tau and another from the hypothetical source VLA~1  \citep{mundt1987}.
But sub-millimeter  \citep{moriarty} and various infrared observations of this region did not confirm the existence of any source in the position of the VLA~1. Presently the HH~151 jet is considered  as a single flow from HL~Tau itself. In this case one should assume that the abrupt changes in its morphology and direction can be produced by variations of the conditions in the surrounding medium.

On the other hand, the neighbouring pre-main-sequence (PMS) star \object{XZ~Tau} is surrounded by an elliptical shell of dimensions $1.5\arcmin \times  2\arcmin$, revealed by the high-resolution imaging in \element[][13]{CO}(1-0)  \citep{welch2000}.
Portions of the wall of this bubble can be seen also in
the optical. Besides, XZ~Tau drives a wide HH outflow \citep{krist1999}. The relation between this flow and the bubble around XZ~Tau is not clear yet, but  we see that HL~Tau  is situated
at the edge of an expanding shell of gas and dust, probably driven by the quasi-spherical wind from XZ~Tau.

Another interesting feature of the HL~Tau jet is that its knotty structure (typical for nearly all  jets) also becomes prominent only after the bending and brightening described above. The Fabry-P\'erot (FP) interferometry of these knots revealed their complex kinematical structure \citep{movsessian2007}, with low and high radial velocity  components of distinctly different morphology. In addition, these knots also demonstrate perceptible proper motions, which were first studied by \citet{mundt90}. Other  data, presented in the papers of \cite{movsessian2007} and \citet {anglada}, showed a certain difference between the proper motions (PM) of the knots, obtained in H$\alpha$ and [\ion{S}{ii}] emission lines. 

There are various theoretical approaches to explain  the formation of knotty structures in the jets  \citep{reipbally}. Usually either various types of instabilities in the jets -- Kelvin-Helmholtz instabilities in hydrodynamical \citep{micono} or magneto-hydrodynamical \citep{ouyed2003} models,  thermal instabilities \citep{gouveia} or time variations in the velocity of the flow that generate internal working surfaces  \citep{raga}, are considered. To prefer one of these approaches, we need more refined observational data. All previous proper motions (PM) investigations of HH jets were performed  in narrow-band images and represent the motions integrated through the full profile of the  emission line. We attempted for the first time  to measure the PM  in the HL~Tau jet for structures with different radial velocities and distinct morphologies, taking into account the possibility of FP scanning interferometry to obtain the velocity-channel images with very high spectral resolution.

\begin{figure*}
\centerline{\includegraphics[width=18 pc,angle=270]{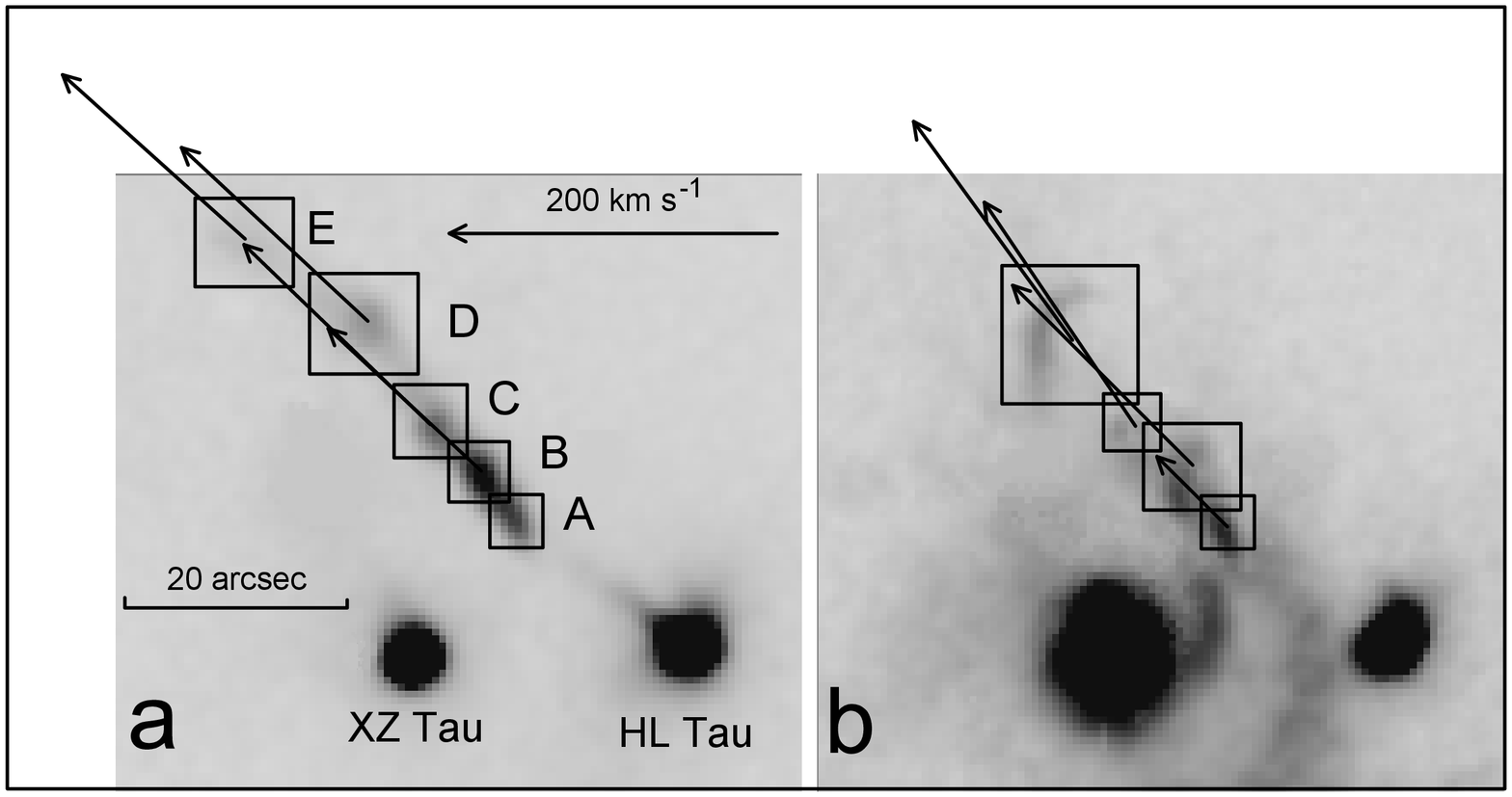}}
\caption{Proper motions of the structures in the A, B, C, D, and E knots of HL~Tau jet, corresponding to high (a) and low (b) radial velocities, are shown by vectors plotted in the images in H$\alpha$ emission corresponding to $-$150 km\,s\,$^{-1}$ and $-$50 km\,s\,$^{-1}$, respectively. The scale of the vectors is indicated by the arrow at the top of the left panel. Rectangles show the areas used for the difference square method.}
\label{fig1}
\end{figure*}

Thus, in this paper we present proper motions of the structures in the HL~Tau jet discussed in our previous paper \citep{movsessian2007}, obtained from  scanning FP observations of this region in  2001 and 2007. We also discuss a scenario in which the formation of tilted bow-shocks and bending of the jet could be the result of interaction of the HL~Tau jet and the XZ~Tau wide flow. This possibility was also proposed in the our previous paper \citep{movsessian2007}.

\section{Observations and data reduction}
\label{observ}

Scanning FP observations in the H$\alpha$ emission line were carried out at the prime focus of the 6 m telescope (SAO, Russia) on 10 February 2007 (seeing was 1 - 1.5\arcsec). We used a Queensgate ET-50 etalon, operating in the 501st order of interference (at the H$\alpha$  wavelength), placed in the collimated beam of the SCORPIO focal reducer \citep{scorpref}. The SCORPIO capabilities for FP interferometry were described by \citet{moisav}. An interference filter with FWHM\,=\,15\AA\ centred on the H$\alpha$ line cut the desired spectral range. The detector was a 2K$\times$2K CCD  EEV 42-40 operated with 2\,$\times$\,2 pixel binning to reduce the read-out time, so 1024\,$\times$\,1024 pixel images were obtained in each spectral channel. The field of view of about 6\arcmin\ was sampled with a scale 0.36\arcsec\ per pixel. The  spectral resolution was estimated as FWHM\,$\approx$\,0.8\AA\ (or $\approx$\,36\,km\,s$^{-1}$) for a range of $\Delta\lambda$\,=\,13\AA\ (or $\approx$\,590\,km\,s$^{-1}$) free from overlapping orders. The number of spectral channels was 36 and the size of a single channel was $\Delta\lambda\approx$\,0.36\AA\
($\approx$\,16\,km\,s\,$^{-1}$).
 
We reduced the interferometric data with an IDL-based software developed at SAO \citep{moisav}. After the primary data reduction, the subtraction of night-sky lines, and wavelength calibration, the observational material consists of ``data cubes''" in which each point in the 1024\,$\times$\,1024 pixel field contains a 36-channel spectrum. 

Proper motions were determined by comparing the  velocity channels, corresponding to   $-$50 km s$^{-1}$ and  $-$150 km s$^{-1}$, obtained in 2001 with the same telescope and optics but another detector \citep[see][]{movsessian2007}, and the new ones of 2007 (all radial velocities used in this paper are heliocentric). To increase the signal/noise (S/N) ratio we summed the three channels in each cube around the above mentioned central velocities. The velocity range in both cases was equal to 31.5 km\,s\,$^{-1}$. Then we determined the PM for both high and low radial velocity structures.

 The first step of a PM measurement from FP data is to reduce the data cubes of both epochs to a common reference system. For the HL~Tau field this procedure has certain difficulties: the reference stars are scarce, and the greater part of these stars are embedded in reflection nebulae. Indeed, we were forced to use nine reference stars, including such nebulous objects as HL~Tau, XZ~Tau and \object{LkH$\alpha$~358}, which definitely diminished the  precision of astrometry. The rms of the differences in the positions of reference stars is about 0.24\arcsec.  
 
The ``difference squared'' method, described in  \citet{hartigan2001} and \citet{reipurth02}, was used. This algorithm is well-adapted for the measurements of extended objects and works   in a selected rectangular region in the images of both epochs, finding the shift in the plane of the sky, which minimizes the sum of squared differences between the intensities in the defined region. For comparison  the proper motions were also determined using the barycenters of the knots. For the HL~Tau jet the systematic differences between both methods were negligible for the compact high-velocity knots, but for the low-velocity bow shocks they became significant. Therefore, only results from the difference square method were used for the proper motion estimates for the bow-shocks.
We estimate the total uncertainty of the position of the correlation peak and astrometry to be about 0.3\arcsec\  for the compact knots and about 0.4\arcsec\  for the extended ones.

\section{Results}
\label{ref}
 As was already described in our previous paper \citep{movsessian2007}, the HL~Tau jet consists of the high-velocity narrow jet with knotty morphology and the low-velocity bows, which are visible only in H$\alpha$ and cross the jet, being located in front of the high-velocity
knots. They exhibit the greater transverse widths at the greater distances from HL~Tau and are not axisymmetric, being  elongated in the SE direction. Their morphology is best visible in HST images \citep{krist2008}.

The knots in HL~Tau jet were described by \citet{mundt1987} and are labelled from A
to E. The connected bow-like structures were detected for knots B, C, and D only. For knot A no visible variations of the morphology in various velocities were detected; the profile of  H$\alpha$   emission in this knot is wide but shows no split into two velocity components. Knot E is visible only in the channels corresponding to high radial velocity. Thus, we estimated the PM of low and high radial velocity structures only in the knots B, C, and D.
However, for the sake of completeness we measured the PM for knot E in the high-velocity range and for knot A using the full line profile.

The results of our PM measurements are presented in Fig.~\ref{fig1}, where the two images of the HL~Tau jet, corresponding to low and high radial velocities, with the vectors of PM of the knots, are shown. The numerical results, again separately  for the low and high radial velocity structures (except for knot A), are given in the Table~\ref{pmtable}.

\begin{table*}
\caption{Proper motions in the jet of HL~Tau}
\label{pmtable}
\centering
\begin{tabular}{l c c c c}
\hline \hline
\\
knot & $\Delta$$\alpha $(arcsec) & $\Delta$$\delta $(arcsec) & V$_{t}$ (km~s$^{-1}$) & P.A. (deg) 
\\
\\
\hline
\\
  &            &     full velocity range       &      &        \\
A &  13.6   & 10.9  &    60 $\pm$ 40 &  45 $\pm$ 19 \\\\
\\
  &   &    high radial velocity structures  &   &       \\
B &  16.4 & 14.7 &   162 $\pm$ 40 &  47 $\pm$ 12 \\  
C &  20.3 & 18.6 &   157 $\pm$ 40 &  46 $\pm$ 15 \\
D &  24.9 & 27.7 &   155 $\pm$ 40 &  47 $\pm$ 19 \\
E &  34.7 & 34.6 &   150 $\pm$ 40 &  47 $\pm$ 19 \\
\\
  &   &    low radial velocity structures &  &     \\
B &  17.1 & 13.7 &   154 $\pm$ 50&  44 $\pm$ 15 \\
C &  19.8 & 17.1 &   167 $\pm$ 50&  35 $\pm$ 24 \\
D &  27.6 & 26.6 &   161 $\pm$ 50&  36 $\pm$ 17 \\

\hline
\newline
\end{tabular}

Column 1: Knot (nomenclature from \citet{mundt90});
Columns 2 and 3:  Right ascension and declination offsets from the position of HL~Tau;
Column 4: Proper motion velocity, assuming a distance of 140 pc; 
Column 5:  Position angle. \end{table*}

The most obvious and interesting result is that the structural components with low and high radial velocity both have very similar values of proper motions. We will discuss this in more detail below. It is also worth mentioning that these new data confirm the tendency of the PM vectors of  low-velocity H$\alpha$ features to turn increasingly north with increasing distance from HL~Tau, found by \cite{movsessian2007}.

\begin{figure}
\centerline{\includegraphics[width=12pc]{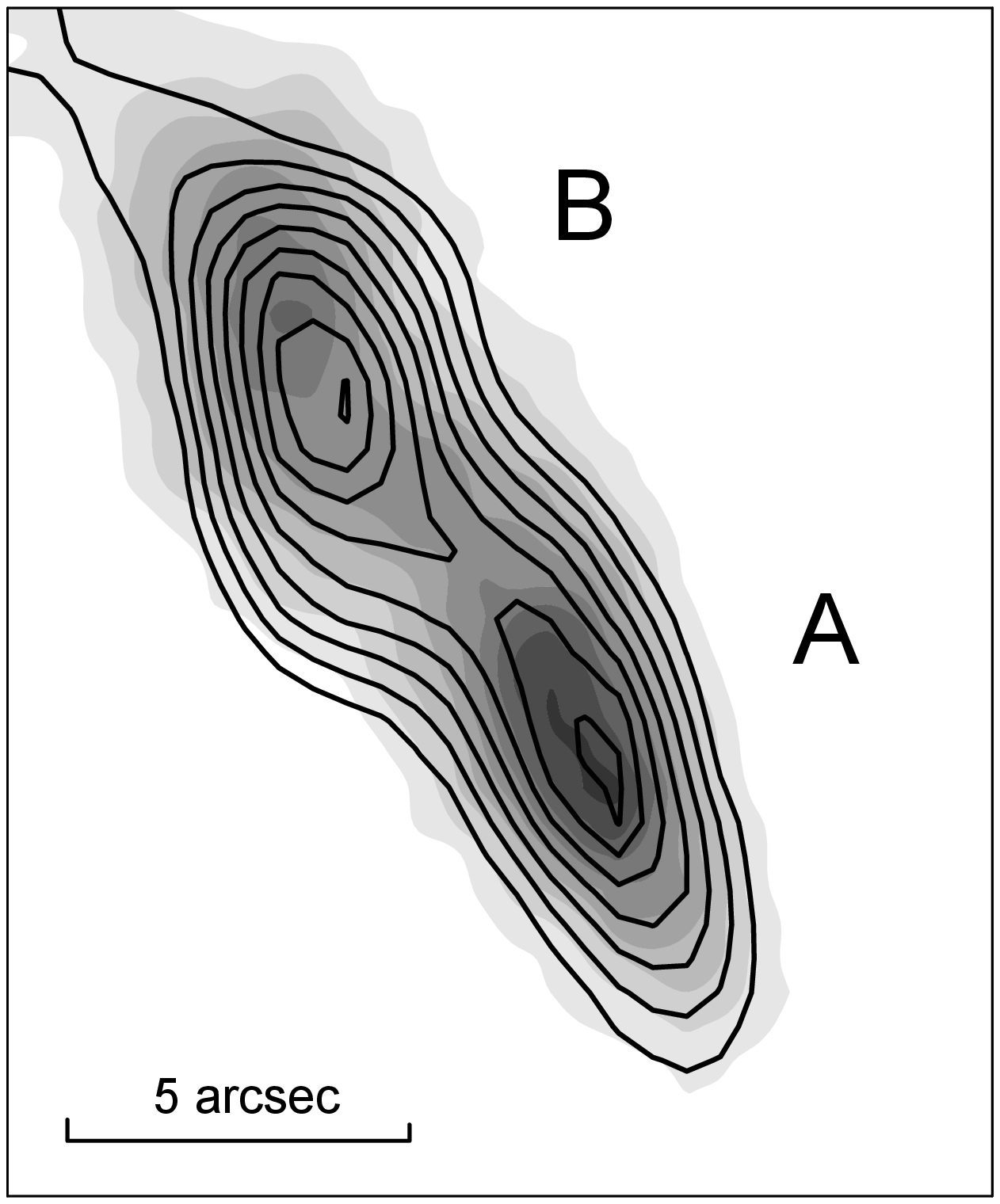}}
\caption{  Appearance of knots A and B in the HL~Tau jet in 2001 (contours) and 2007 (greyscale). The progressive elongation of knot A and the obvious shift of knot B can be compared.}
\label{fig2}
\end{figure}

\begin{figure}
\centerline{\includegraphics[width=14pc,angle=270]{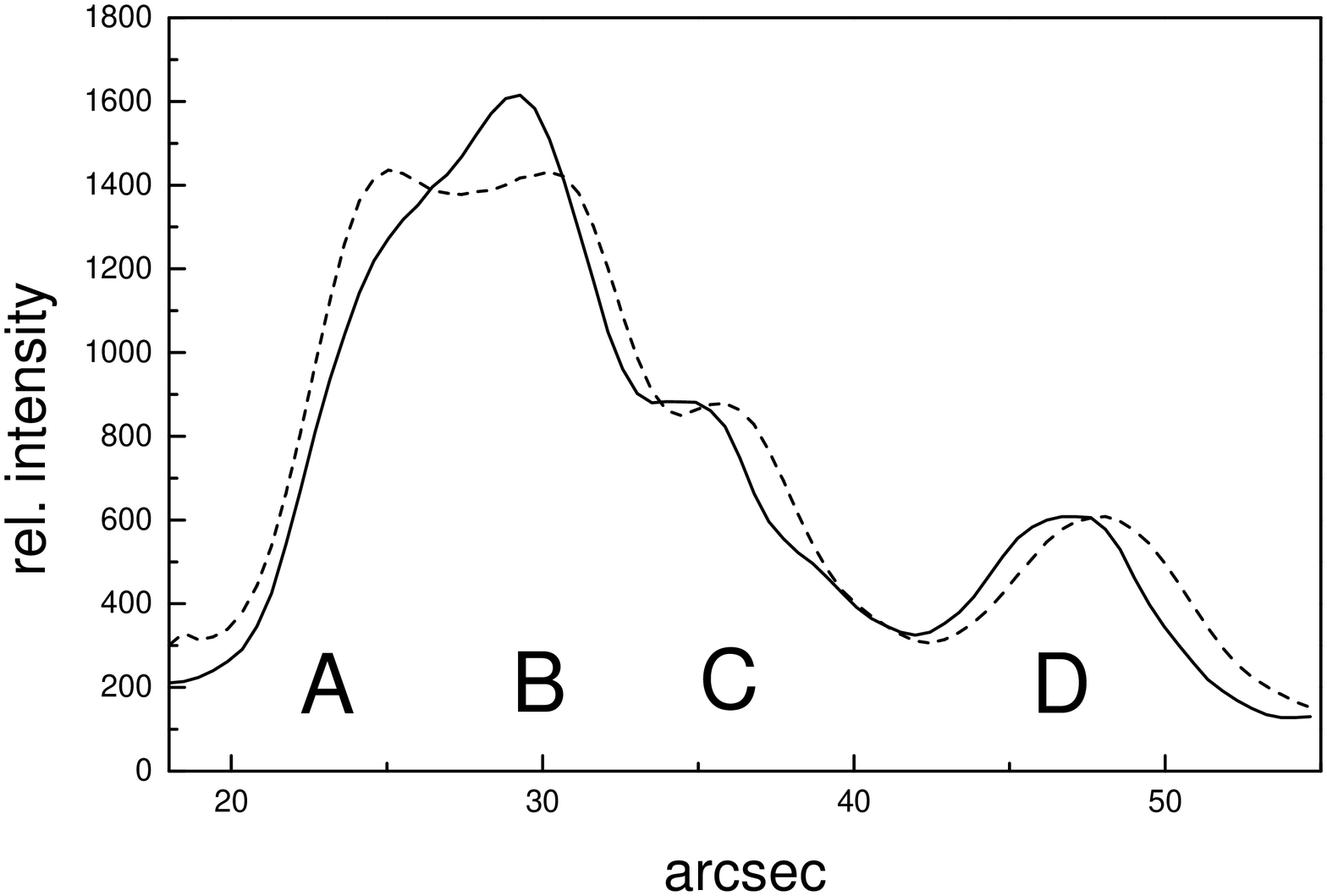}}
\caption{ Cuts along the HL~Tau jet for the two epochs of observations (full line: 2001, dashed line: 2007), with the designations of the knots shown below. The intensities of both   cuts are normalized by knots C and D. Shifts of the knots are apparent except for the beginning of knot A.}
\label{fig3}
\end{figure}

When considering the motions of the knots between two epochs their
shift is obvious except for knot A. Its behaviour can be described as the progressive elongation in the direction of its motion, leaving its trailing edge nearly fixed. This is clearly seen in Fig.~\ref{fig2}, where the images of knots A and B in the two epochs, integrated by full line profile, are presented.\

For further clarification we show in Fig.~\ref{fig3}\  the intensity slice along the HL~Tau jet in integral H$\alpha$ emission. The nomenclature of the knots is the same as in \citet{mundt90}. The shift of the knots between two epoches is apparent except at the beginning of knot A where the abrupt brightening of the flow begins. This point can be considered as stationary relative to HL~Tau. The change of the relative brightness of knots A and B between 2001 and 2007 should also be noted. 

 Comparing the present measurements with the previous ones, we see that the directions are quite similar throughout the literature, though the values differ significantly. The present PM values  are more similar to the results of \citet{anglada} than to the higher estimates of \citet{mundt90} and \citet{movsessian2007}. Very probably this discrepancy results from the unavoidable systematic errors of various origin, which were briefly discussed also in Sect.\ref{observ}. In any case, this inconsistency cannot affect the basic results and the subsequent  discussion in any way.

\section{Discussion}
\label{disc}
\subsection{Multiple working surfaces in the HL~Tau jet}

The complex kinematical picture of knots in the HL~Tau jet speaks about the existence of two distinct shock structures in each knot. These structures differ not only by their radial velocity but also by excitation, because bow-shape structures are visible only in H$\alpha$ emission and high-velocity knots are much brighter in [\ion{S}{ii}] \citep{movsessian2007,krist2008}. This picture strongly resembles separation by excitation of bow-shock and Mach disk structures in the terminal working surfaces of the jets \citep{raga}. This division of shocked regions was indeed was found through narrow-band imagery \citep{reipurth92}. Moreover, with the usage of  spectral imagery, these regions were also separated kinematically \citep{hartigan2001, morse1992, movsessian2000}.  But, for the jet knots even the doubtless splitting  of two shocked structures -- both by radial velocity and by excitation --  leaves some uncertainty in choosing between the theoretical mechanisms of the knot formation.
Here, in our opinion, the proper motion values combined with the radial velocities can play the crucial role in clarifying this situation. 

As we already showed, the proper motions of the two distinct radial velocity structures in each knot of the HL~Tau jet  have the same values. To analyse this picture we suggest considering the knots as the internal working surfaces formed in a time-dependent flow, i.e. in the same way as the terminal working surfaces. In this case they should contain the two principal shock regions: the ``reverse shock'' (or Mach disk) that decelerates the supersonic flow, and the ``forward shock'' (or bow-shock) that accelerates the ambient medium. The pattern speed of these two structures should be the same, in contrast to the  fluid speed of emitting particles.

Indeed, our data allow us to obtain for each knot both the fluid speed, which corresponds to the velocity of emitting particles, and the pattern speed of the visible structures. 
Tangential velocity is the projection of  pattern velocity
onto the plane of the sky: V$_{t}$ = V$_{pattern}~\cos{\alpha}$, where $\alpha$ is the angle between the outflow axis and the plane of the sky. Accordingly, the radial velocity of the knot corresponds to the projection of the velocity of emitting particles (fluid speed) on the line of sight: V$_{r}$ = V$_{fluid}~\sin{\alpha}$.

The ratio $\zeta$  of pattern speed to fluid speed was for the first time determined for \object{HH~34} jet by \cite{EisloffelMundt}, but the kinematical structures in the knots were not separated in their work; only the spectral imagery technique allows this. 

To obtain the correct estimates for the jet absolute velocity and inclination ($\alpha$ angle) we need to use the structures where  $\zeta$  = 1.  For example, in a Mach disk by definition the fluid and pattern velocities have nearly the same values, so  $\zeta$   should be near unity. Thus, combining the tangential and radial velocities of  high-velocity compact knots (which we assume to represent Mach disks), we find that $\alpha$ = 50$\degr$ $\pm$ 10$\degr$ and the mean value of the absolute velocity of the jet is about 250 km s$^{-1}$.  

    Previous observational estimates for $\alpha$ put  its value  in the range 40$\degr$-60$\degr$ \citep{anglada,movsessian2007}. On the other hand, models of the HL~Tau circumstellar dusty environment provide an angle of the flow about 43$\degr$ \citep{men'shchikov}. As one can see, the new values agree well with the previous estimates.

Now we can consider the low radial velocity, bow-shape structures, which we assume to be the shock waves that entrain particles and rise the temperature  of the ambient medium. Knowing the inclination angle, we obtain the mean value $\zeta$=2.6 for the three bows in the HL~Tau jet, which strongly differs from the compact knots. 

 As was shown by \citet{hartigan1989}, the  ratio of radial velocities in shocked structures depends on the flow and ambient medium densities as well as on the jet velocity itself. From the same model, it follows that the relative surface brightness of
 a Mach disk and a bow-shock also depends on the ratio of the  jet and ambient medium densities (n$_{j}$ and n$_{a}$ correspondingly), and brighter bow-shocks are formed in denser jets. For knots within the HL~Tau jet,  the relative surface brightness of bow-shock and Mach disk is nearly equal, so the ratio n$_{a}$/n$_{j}$ is a little more than unity. Thus, the sudden increase of  brightness of the jet and appearance of bright bow-shocks, starting from the
20$\arcsec$ distance from the source, could be understood as the result of an abrupt change of the physical conditions in the ambient medium,  when the flow reaches the region of lower density formed by the poorly collimated outflow from XZ~Tau.

Of course, one cannot exclude the possibility that this brightening can be the result of the lower line-of-sight extinction after the deflection point. However, both in our images and in Fig.4 from the paper of \cite{krist2008} we can see that the narrow jet is visible from the star and almost to the deflection point; no H$\alpha$ emission low-velocity structures exist in this range.

\subsection{Jet - wind interaction?}

Now we can consider the unusual behaviour of  the HL~Tau jet, gathering all its unusual features. Some of them were already discussed in our first paper \citep {movsessian2007}. In summary, we have
the following points:\begin{itemize}

\renewcommand{\labelitemi}{$\bullet$}

\item The bend of jet at the distance of about 20\arcsec\ from the source star;

\item the abrupt increase in the brightness of the jet, starting from the same position;

\item position of this point is not shifted between the two epochs of observation;

\item appearance of low radial velocity structures only after the bending of the jet;
\item  these structures have the shape of non-axisymmetric bows with brighter SE edges;
\item transverse velocity gradient is observed in these low-velocity components.
\end{itemize}

We suppose that these phenomena are mutually interrelated and, as was mentioned above, that they can result from the abrupt change in the physical condition.

        Deflection of the HL~Tau jet to about 12$\degr$ could occur from the glancing collision between the jet and the cloud material, located approximately
at 20\arcsec\ distance from HL~Tau \citep{krist2008}, as in case of \object{HH~110,} where the jet collides with a dense clump (Reipurth et al. 1996). But, as was mentioned above, the relative brightening of bow-shocks (which is  observed in this case) takes place when the flow reaches the region with lower ambient density. 

\begin{figure}
\centerline{\includegraphics[width=17pc]{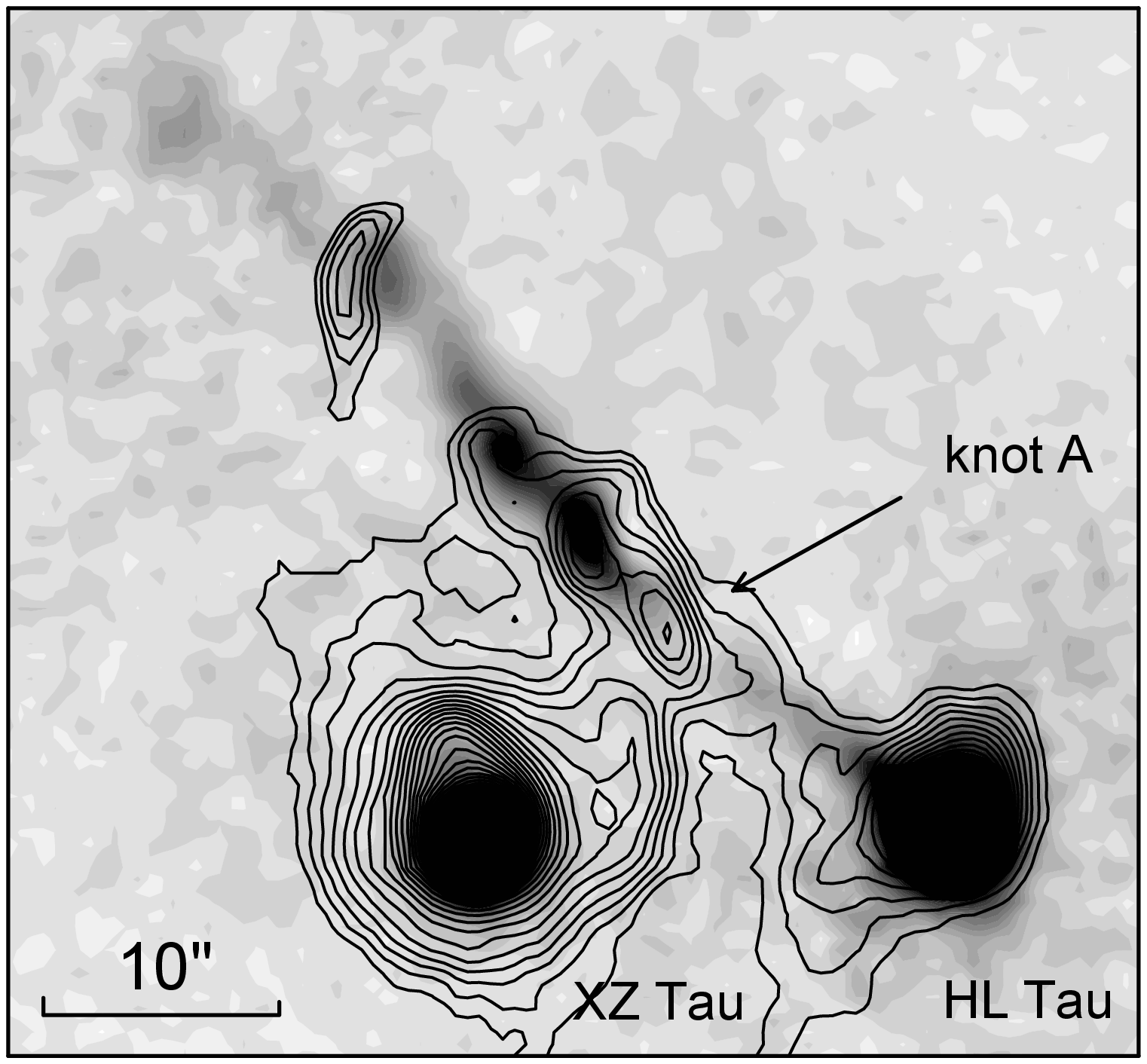}}
\caption{ Superposition of  the HL/XZ~Tau FP images in the monochromatic [\ion{S}{ii}] $\lambda$6716\AA\  line emission (greyscale) and low-velocity H$\alpha$ emission (contours).  The H$\alpha$ observations were performed in Sept 2001 (see Sect. \ref{observ}), the   [\ion{S}{ii}] data were taken in Jan 2003 \citep{movsessian2007}. Knot A (shown by arrow) is visible only in H$\alpha$ and appears as a continuation of the H$\alpha$ emission arcuate structure west of XZ~Tau.}
\label{fig4}
\end{figure}

In this connection it is important to consider again knot A. Unlike the other knots, its edge, nearest to the star,  has no
perceptible proper motion (Figs.~\ref{fig2} and~\ref{fig3}); on the other hand, the knot is detectable only in H$\alpha$ emission (Fig.~\ref{fig4}).  We are inclined to believe that the tail of knot A represents a so-called deflection shock, formed at the dense wall created by the wind from XZ~Tau. Both its stationarity and morphology have a great similarity to the  feature that was recently discovered in the \object{HH~46/47} jet by \cite{hartigan2011} (compare their Fig.14 and
the  Fig.4 from the paper of \cite{krist2008}) and considered by them as the same kind of deflection shock.

On the other hand, the observed bending of the jet can be produced by the ram pressure of a side wind, as was shown in \citet{ciardi}. An expanding shell around XZ~Tau, detected by \citet{welch2000}, can definitely   play the role of such a side wind. As a result of this interaction, the increase of shock energy release is manifested by the brightening of the jet. The formation of elongated wings of the bows in the direction of the XZ~Tau bubble (to the SE) as well as the transverse velocity gradient detected in bow-shape structures support this idea. In any case, the transverse velocity gradient is too large to be explained by jet rotation.

One cannot exclude the possibility that both mechanisms can play role in the observed behaviour of the HL Tau jet. Using the model described by \cite{cantoraga} we estimated, however, that the expansion velocity of XZ Tau bubble \citep{welch2000} is insufficient for the observed abrupt bend of the HL Tau jet. But, as was already mentioned above, it can be responsible for brightening of the jet.

We also recall that according to our previous measurements \citep {movsessian2007},
confirmed also by other authors, in the SW direction from XZ~Tau  the definite turn in the PM vectors of the knots in the so-called ``H$\alpha$ jet'' is observed. As was also suggested by \citet{movsessian2007}, this can be caused by the interaction between this flow and the south-western side of XZ~Tau bubble. These two cases of the unusual behaviour of the jets around the XZ~Tau, in our opinion, argue that a real interaction  is taking place between
the jets and the expanding bubble of this star.

\section{Conclusion}
\label{concl}

  For the first time the proper motions of the knots in the collimated\  jet were measured with the scanning Fabry-Per\'ot interferometer.  This allows us to make certain suggestions and conclusions. In the end we summarize our findings in the following way:

(1) Our PM measurements  show that the two structures distinguishable by radial velocity that exist in the HL~Tau jet  have the same tangential velocities, with a typical value of the order of 160 km s$^{-1}$. 

(2) The angle of the flow to the sky plane,
 determined from the radial and tangential velocities
of high-velocity structures,
is  50$\degr$ and the total space velocity of the flow is about 250 km s$^{-1}$.

(3) We are inclined to assume that the different velocity structures in each knot of the HL~Tau jet are detecting two principal shocks,  namely bow-shock and Mach disk.  

(4)
Taking into account all observed features, we conclude that the knots in the jet originate from the episodic  velocity variations  in the HL~Tau outflow, and each of them forms internal working surfaces with two principal shock regions. 

(5)
The strange and unusual behaviour of  the HL~Tau jet, starting from $\sim$20$\arcsec$ from the source, is
connected with the abrupt change of physical conditions in the ambient medium.
Namely, the jet enters the zone where the matter is blown out by the  wide wind from XZ~Tau, creating a deflection shock on its border where the swept-out matter   formed the dense shell. 

 Recent narrow-band observations of the knots in HH~34 by \cite{hartigan2011} demonstrated
that those can also be resolved to the high- and low-excitation regions, and that in each knot both shock regions have the same proper motions.
This interesting result can be considered to support our interpretation of the high and low radial velocity structures in the HL Tau jet. 

We hope that this work, together with other new results, will give an impulse to develop new theoretical models of the collimated outflows, which will take into consideration all modern observational data.


\begin{acknowledgements}
 The authors thank Bo Reipurth and John Bally for reading the preliminary version of the paper and for helpful discussions. Authors also express their thanks to the referee, whose comments and critical suggestions helped to improve the paper. This work was partially supported by ANSEF grant
PS-astroex-2761.
This work was performed partly during the stay in MPIA (T.A.M.) in Heidelberg supported by DAAD scientific exchange program.

\end{acknowledgements}

\bibliographystyle{aa}

{}
\end{document}